# Uniformly Distributed $Fe_2O_3$ Nanoparticles' Thin Films Synthesized by Spray Pyrolysis

Dayanand[1,$], Meenu Chahar[1,$], Devesh Kumar Pathak[2,$], O. P.Thakur[*1], V. D. Vankar[1] and Rajesh Kumar[2,*]

[1]Material Analysis and Research Laboratory, Department of Physics, Netaji Subhas University of Technology, New Delhi- 110078, India

[2] Materials and Device Laboratory, Discipline of Physics, Indian Institute of Technology Indore, Simrol, Indore-453552, India

[$] Authors contributed equally

*Email: rajeshkumar@iiti.ac.in (Rajesh Kumar); opthakur@nsut.ac.in (O.P. Thakur).

**Abstract**

Thin films of uniformly distributed $Fe_2O_3$ nanoparticles have been prepared on single crystal silicon and glass substrates by a spray pyrolysis technique in a single step using a mixture of water and ferrocene dissolved in xylene. The size distribution of nanoparticles is found to be in the range of 20 nm to 30 nm. The films have been characterized by X-ray diffraction, scanning electron microscopy and Raman spectroscopy techniques. The uniformity of the grown film was evident from the electron microscopic images on both the substrates. The crystallinity and band gap were investigated using X-ray diffraction and absorption spectroscopy respectively. Raman measurements of the prepared films have been carried out using two excitation wavelengths of 633 nm and 785 nm to investigate the depth of homogeneity of the films. The wavelength dependent Raman measurements reveal that the film is uniform across the thickness of the film on both the substrates.

**Keywords:** *Spray pyrolysis; Ferrocene; Hematite; Nanoparticles; Thin films.*

# Introduction

In recent years, iron oxide ($Fe_2O_3$) has attracted much attention as its different structural and morphological forms have found several technological applications such as magnetic storag devices, magnetic fluid devices, pigments, catalysis, electromagnetic shielding etc[1–5]. Iron oxide exists in different structural forms viz- rhombohedral $\alpha$- $Fe_2O_3$ (hematite JCPDS-card No. 89-0597), cubic $\alpha$–$Fe_2O_3$ (maghemite JCPDS- card No.39-1346) and $Fe_3O_4$ (magnetite JCPDS card No.19-0629). These structural forms exist in different temperature ranges and exhibit different physical and chemical properties. These materials can be synthesized by variety of techniques using different precursors and have variety of applications [6–17]. Some of the technological applications and synthesis processes used by different workers are given in the table no.3 below. It is observed that very fine powders, or hollow spheres[5,10] can be grown by hydro-thermal process[7,18–23] or by spray pyrolysis[14] in large amounts whereas good quality oxide films having great adhesion to the substrate can be prepated using electro-deposition methods[24]. The co- precipitation process is used to obtain core-shell nano-particles[3]. Heat treatment of iron substrates leads to growth of large arrays of vertically aligned nano-rods of $Fe_2O_3$ [11]. These particles do have different orders of crystallinity[12,13,15]. The kinetics of growth of some of these different structures has been studied by Jorgensen et.al[8]. Since, the morphology, microstructures and the crystallinity are critical factors to bring novel and unique size and shape dependent functional properties, significant attention has been paid for controlled synthesis of nano- materials employing different techniques and variety of pre-cursers. Most of these techniques involve cumbersome routes and complex chemical reactions of the pre- cursers involved. Recently Gonzalez- Carreno et al.[15] have used spray pyrolysis technique to prepare uniform sized $\gamma$- $Fe_2O_3$ particles of different degree of crystallinity using different precursors such as Fe-acetylacetone, Fe-ammonium citrate, Fe- nitrate and Fe-chloride dissolved in methanol. They have also shown that $\alpha$-$Fe_2O_3$ can be prepared in a single step by combustion of iron nitrate and malonic acid dihydrazide by spray pyrolysis technique. Pressure pyrolysis technique has been used by Hirano et al.[16] for synthesis of ferrite- carbon composites using nickelocene and vinylferrocene as pre-cursers. This technique, however, employs very high pressures (30-200 MPa) and high temperatures (500- 700°C) in a hydrothermal growth apparatus. Spray pyrolysis is a widely used technique for producing micro-and nano-scale particles of varying size distributions from materials of various kind such as metals, oxides semiconductors etc. Because of its convenient process characteristics compared to other methods like physical and chemical vapour deposition, plasma assisted vapor deposition, laser ablation etc., spray pyrolysis has found tremendous attention in recent times. Spray pyrolysis is a simple and low cost technique capable of providing large area high quality adherent films of uniform thickness. In this process, simple pre-cursers could be used to obtain fine particles or uniform coatings over large area in a single step avoiding complicated processes and complex chemical reagents.

Basic technique involved in the spray pyrolysis process is the pyrolytic decomposition of salts of a desired compound to be deposited. The spray process involves atomization of the aqueous (or solvent) pre-cursor solution into a spray of fine droplets through a fine nozzle. These droplets undergo volume expansion during their flight from the nozzle towards the hot substrate due to a temperature gradient between the two. As the droplets arrive on the hot substrate they further expand and burst into finer droplets which undergo pyrolytic decomposition and pulverization into nano-sized crystallites[17]. The substrates provide thermal energy for the thermal decomposition and subsequent recombination of the constituent species followed by sintering, nucleation, growth and crystallization, resulting in a uniform coherent and continuous thin film (Figure1a and1b).

In this paper, a single step spray pyrolysis method for preparation of thin films of $\alpha$-$Fe_2O_3$, consisting of uniformly distributed nano- particles is reported. The films have been deposited on silicon and glass substrates by using a solution of ferrocene dissolved in xylene mixed with water. The phase identification and microstructure analysis has been carried out by X-ray diffraction (XRD), Raman spectroscopy and scanning electron microscopic (SEM) techniques. Raman spectroscopy has been carried out using two excitation wavelengths to examine the homogeneity along the film depth. The α-Fe2O3 film deposited on glass substrate was used to estimate the band gap of the film using absorption spectroscopy.

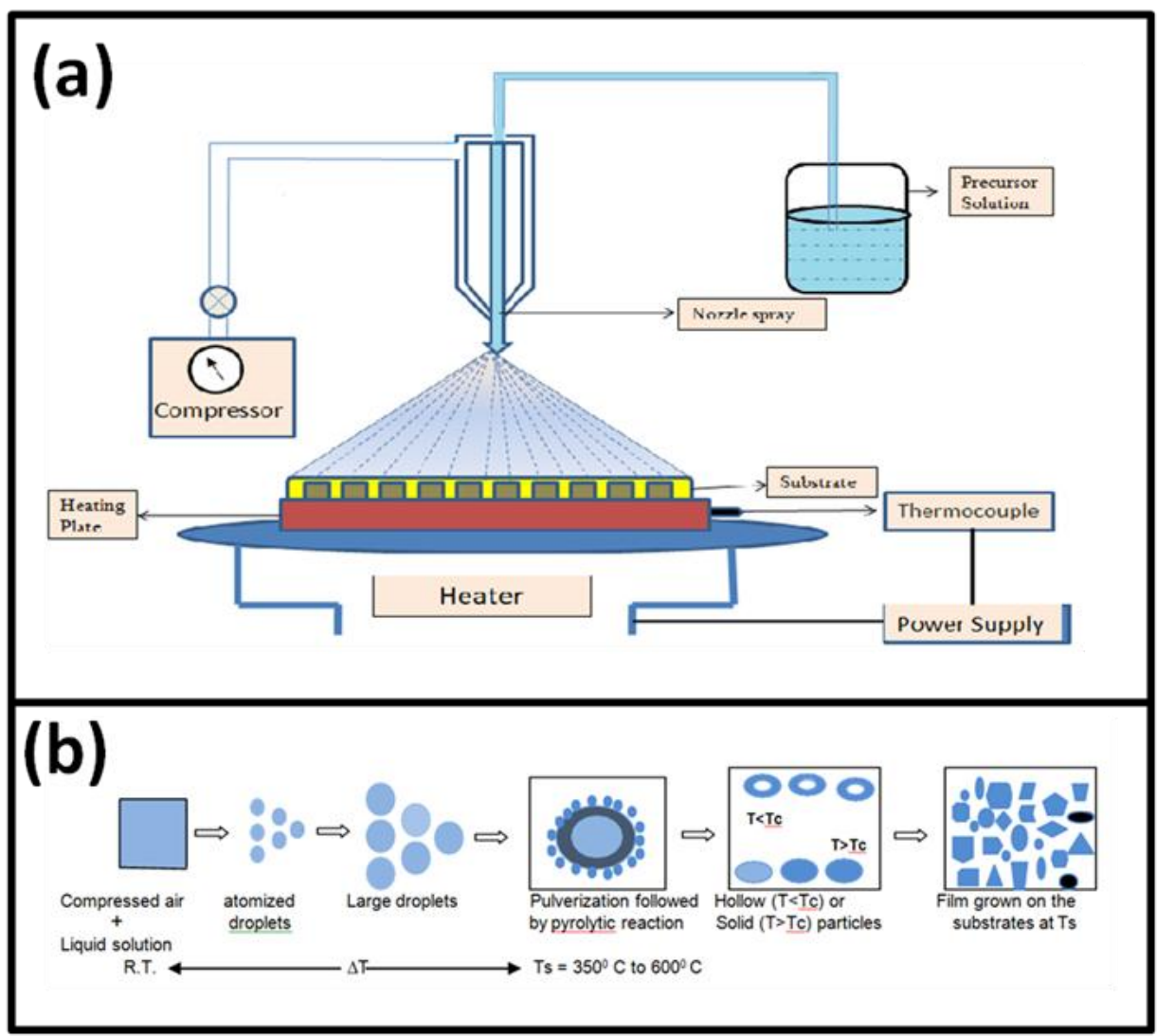

Figure 1: (a) Schematic diagram of the spray pyrolysis set up used for deposition of uniform thin films over large area. It consists of a shielded heater plate, heat shield plate, temperature controller, liquid dispenser attached to a spray nozzle- atomizer and an air compressor. (b) A schematic representation of the mechanism of nano-particle formation by the spray pyrolysis process.

**Experimental Details**

The films of α-$Fe_2O_3$ were prepared by spray pyrolysis by an apparatus shown in Figure(1a). Ferrocene was dissolved in xylene (solubility ~ 13%) and equal amount of pure water was added into this solution to form a suspension. The suspension (composition : 2.5 gms. of ferrocene dissolved in 125 mL of xylene and 125 mL of pure water ) was continuously stirred and fed into the spray nozzle (orifice diameter ~0.2 mm) and atomizer system to generate very fine droplets which were directed on a number of heated substrates kept on a hot plate maintained at constant temperature. The nozzle to substrate distance was ~25 cm forming a cone of diameter 20 cm on the substrate plate. The solution was sprayed for ~ 20 minutes. Thin films were deposited on both glass (1 mm thick) and silicon single crystal wafers (300 μm thick) kept on the same heater plate. The temperature of the substrates was monitored by using a chromel −alumel thermocouple. The measured temperature on the silicon substrate was 500 °C and that on the glass substrate surface was 480 °C. After the deposition process, the heater was switched off and allowed to cool till room temperature was achieved. The time needed was ~ 2 hours. The films were annealed at gradually decreasing temperatures under ambient conditions. The films (~100 nm thick) were then taken out for SEM, XRD, UV-Visible and Raman spectroscopy analysis.

The thin films of $Fe_2O_3$ on silicon as well as glass substrates were examined under a Zeiss FE SEM model Supra Z5 in the secondary electron mode with excitation voltage 5 kV and working distance of 3.3 mm. The X-ray diffraction data was recorded using Rigaku Smart Lab monochromatic Cu-Kα radiation (λ=1.54 Å) in glancing angle diffraction mode keeping the incidence angle α = 2° (Figure 2). The Raman spectra were recorded using Horiba Labram HR evolution laser He-Ne laser ($λ_{ex}$= 633 nm and 785 nm).

## Results and Discussion

Figure 2a and 2b show typical scanning electron micrographs of these samples on silicon and glass substrates respectively. The micrograph (Figure 2a) shows uniform and dense distribution of small clusters consisting of particle of random shape and size between 20 nm to 30 nm. The micrograph in Figure 2(b) shows uniformly distributed larger clusters consisting of particles of random shape and size between 20 nm to 50 nm. These particles are not so densely distributed as compared to those deposited on silicon substrates (Figure 2a). The difference in size and distribution is attributed to difference in the substrate temperature at which the pyrolysis was carried out viz at 500° C on silicon and 480 °C on glass substrates.

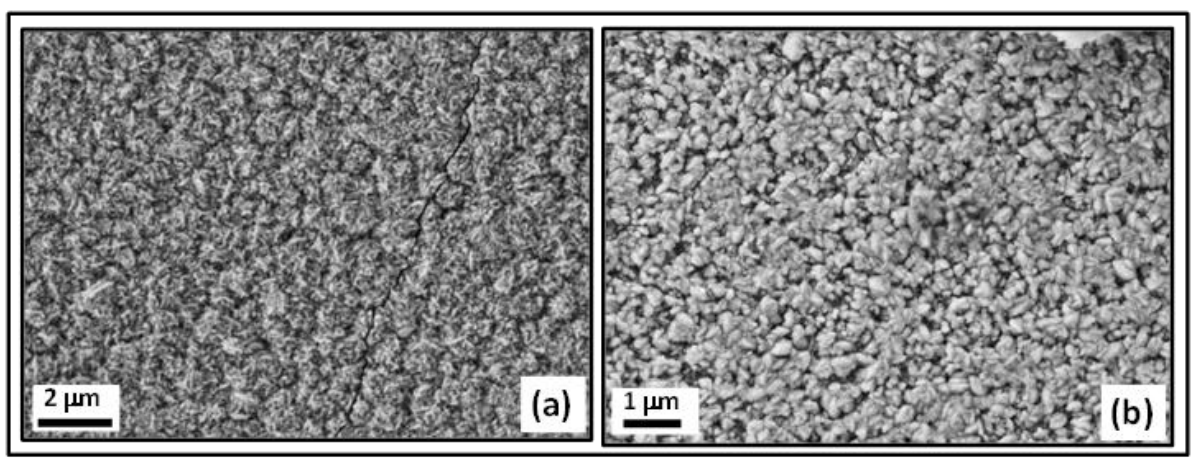

Figure 2: SEM micrographs of α-$Fe_2O_3$ thin films deposited on (a) silicon single crystal substrate at 500 °C and (b) glass substrate at 480°C.

Figure 3 shows XRD patterns of the films deposited on silicon and glass substrates respectively. All the peaks could be identified with the hexagonal phase of α-$Fe_2O_3$ with structural parameters of a = b = 5.034 Å and c = 13.721 Å, which are in good agreement with the literature ( JCPDS card no. 33-0664). It is reported that pure α-$Fe_2O_3$ phase exists in rhombohedral form (JCPDS No. 89-0597) with a small amount of carbon in graphitic form. The detail of analysis is presented in Table-1. Since in our synthesis process both carbon and oxygen could be present in trace amounts they could be responsible for stabilizing the hexagonal phase with slightly increased lattice parameter values as compared to those reported in the literature. XRD patterns do not show any evidence of carbon present in the films. Some of the peaks are identified as due to silicon which is coming from the single crystal silicon substrate.

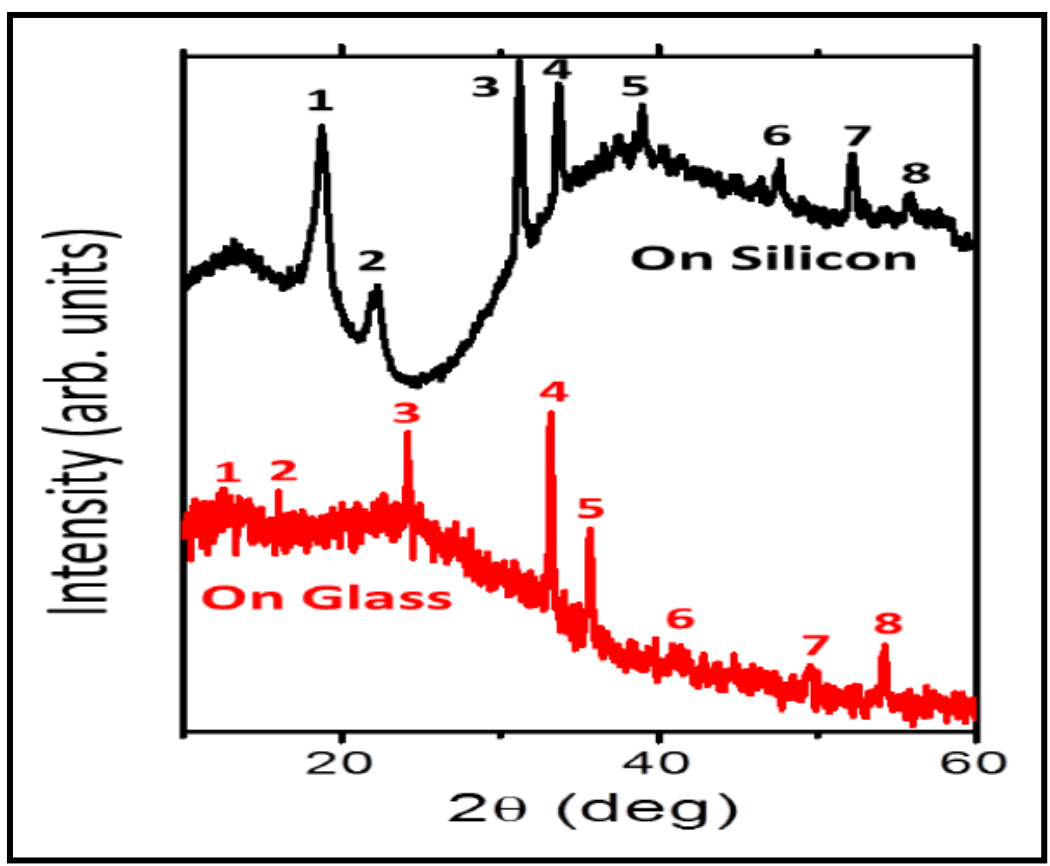

Figure 3: X-ray diffraction patterns of films deposited on (a) silicon substrate and (b) glass substrate. The data was recorded using Cu-Kα radiation in glancing angle diffraction mode keeping incidence angle α =2°.

Figure4 shows the Raman spectra from the deposited films on both the substrates recorded using two excitation wavelengths of 633 nm (left panel) and 785 nm (right panel). The peaks observed in Raman spectra in Figure 4(a) and 4(b) have been compared with the data in the literature and-

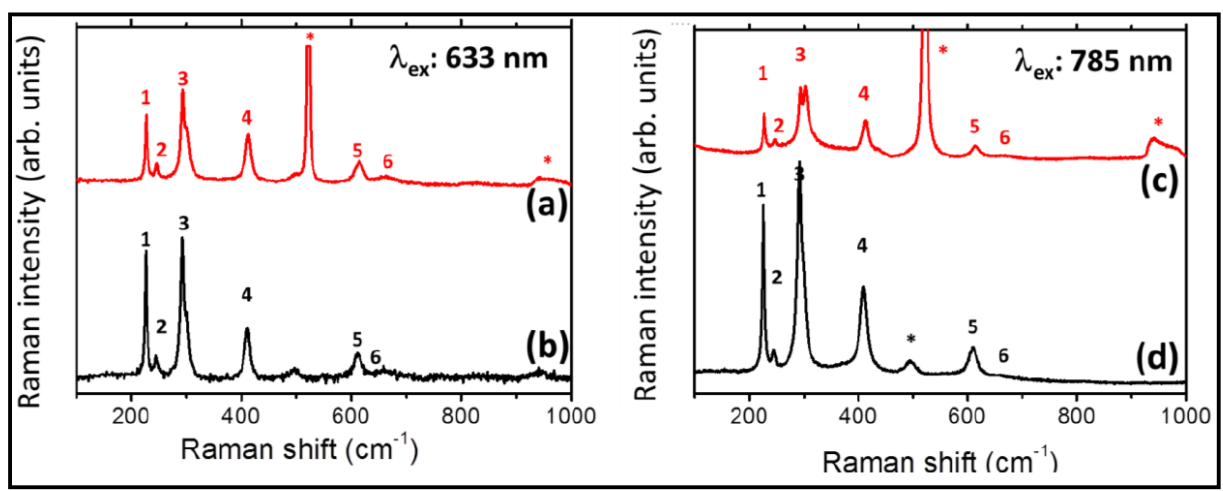

Figure 4: Raman spectra, recorded using excitation wavelength of 633 nm (left panel) and 785 nm (right panel), of films deposited on silicon substrates (a,c) and on glass substrates (b,d).

-summarized in Table 2. The films on both the substrates show Raman peaks corresponding to two A1g and five $E_g$ modes belonging to α-$Fe_2O_3$. No evidence of carbon could be identified in these films. The Raman peaks corresponding to the substrates have been marked with (*). A careful analysis at the Raman data can be carried out to obtained more information related to the homogeneity of the sample. A comparison between the Raman spectra in Figure 4a with 4c and Figure 4b with 4d reveals that Raman modes are identical and has no effect of excitation wavelength. It is important here to mention that the two excitation wavelengths will have different penetration depths in the sample and thus the spectra contain information up to this depth. Considering that, excitation wavelength immune Raman signal means the presence of same material up to the sample depth penetrated by the excitation wavelength. It means that the α-$Fe_2O_3$ film deposited on either of the substrates is uniform along the depth. The band gap of α-$Fe_2O_3$ film deposited on the glass substrate was calculated based on its UV-Vis absorption spectrum (Figure 5) and estimated to be 3.24 eV. A high band gap value of the deposited film indicates the insulating nature of the film deposited on glass substrate.

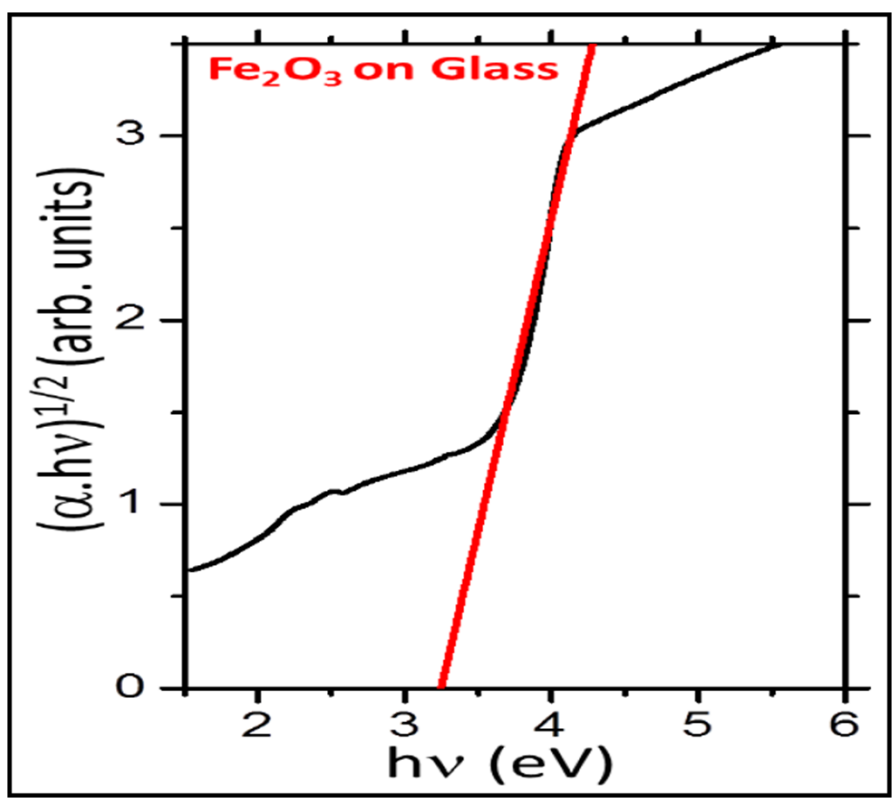

Figure 5: UV-Vis spectrum (black curve) of film deposited on glass where the red line shows the straight line fit used to estimate the band gap.

A comparison of the structural, crystallographic and optical spectroscopic (UV-Vis and Raman) data reveals that a uniform film, consisting of α-$Fe_2O_3$ nanostructures, can be prepared using spray pyrolysis. The film, so prepared, has a band gap of 3.24 eV and is homogeneous along the depth of the film

## Conclusions

Spray pyrolysis has successfully been used for deposition of dense thin films of uniformly distributed α-$Fe_2O_3$ nano-particles on crystalline as well as amorphous substrates as revealed from the morphological studies. The single step spray pyrolysis process reported here uses a mixture of water and ferrocene dissolved in xylene. The structural analysis carried out using XRD, shows that $Fe_2O_3$ nano-particles have a single phase hematite structure stabilized with trace amount of carbon. Optical absorption spectroscopy was used to estimate the band gap of the deposited film on glass substrate which yields band gap of 3.24 eV indicating an insulating behavior. An excitation wavelength dependent Raman spectroscopy was used to check planar and depth homogeneity and to validate phases which were found consistent with the XRD analysis. In brief, the comparison of structural, morphological and spectroscopic results concludes that a uniform film, consisting of $\alpha$-$Fe_2O_3$ nanostructures, can be prepared using spray pyrolysis. The films, so prepared, have a relatively high band gap and are homogeneous along the depth of the films.

## Acknowledgements

The authors are thankful to the Vice Chancellor, Netaji Subhas University of Technology, New Delhi for financial support. Authors acknowledge SIC facility (IIT Indore) for various measurements. One of the authors (D.K.P.) acknowledges Council of Scientific and Industrial Research (CSIR) for financial assistance (File No. 09/269 1022(0039)/2017-EMR-I). Facilities received from Department of Science and Technology (DST), Government of India, under FIST scheme (Grant SR/FST/PSI-225/2016) is highly acknowledged. Authors acknowledges financial support from Science and Engineering Research Board, Govt. of India (Grant no. CRG/2019/000371).

**Table 1: Analysis of XRD Pattern for $Fe_2O_3$ deposited on silicon and glass substrates**

| Sl. No. | $2\theta$ values on **Silicon Substrate** (observed) | Sl. No. | $2\theta$ values on **Glass Substrate** (observed) | Assigned hkl values of $Fe_2O_3$ (JCPDF) | Assigned hkl values of **Silicon** |
|---|---|---|---|---|---|
| 1 | 18.909 | 1 | 12.364 | -- | -- |
| 2 | 22.303 | 2 | 16.026 | -- | -- |
| 3 | 31.333 | 3 | 24.122 | 24.138 (012) | -- |
| 4 | 33.848 | 4 | 33.265 | 33.153 (104) | 28.444 (111) |
| 5 | 39.000 | 5 | 35.660 | 35.612 (110) | -- |
| 6 | 47.666 | 6 | 41.250 | 40.855 (113) | -- |
| 7 | 52.181 | 7 | 49.815 | 49.480 (024) | 47.306 (220) |
| 8 | 55.969 | 8 | 54.386 | 54.091 (116) | -- |
|  | -- |  | -- | 57.429 (122) | 56.126 (311) |
|  | -- |  | -- | 57.590 (018) | -- |

**Table 2: Analysis of Raman Spectra for $Fe_2O_3$ deposited on Silicon and Glass substrates recorded using excitation wavelength of 633 nm.**

| Sl. No. | Observed Raman Peak positions on Silicon Substrate ($cm^{-1}$) | Sl. No. | Observed Peak positions on Glass Substrate ($cm^{-1}$) | From Ref. 29 ($cm^{-1}$) | From Ref. 25 ($cm^{-1}$) | Mode Assignment Ref. 29 & 25 |
|---|---|---|---|---|---|---|
| 1 | 225 | 1 | 225 | 226.5 | 223 | A1g |
| 2 | 249.3 | 2 | 241.5 | 245.5 | | Eg |
| 3 | 294.1 | 3 | 294.1 | 293.5 | 291 | Eg |
| 4 | 414.9 | 4 | 411 | 413 | 409 | Eg |
| 5 | 612.6 | 5 | 612.6 | 612.5 | 596 | Eg |
| 6 | 661.3 | 6 | 661.3 | 659 | | Disorder (?) |

**Table 3: Synthesis of different structural /morphological forms of $Fe_2O_3$ and their applications:**

| Sl. No. | Structure/ Morphological form | Synthesis method | Pre-cursers used | Property | Application | Reference |
|---|---|---|---|---|---|---|
| 1 | Ultra fine $Fe_2O_3$ and $Fe_3O_4$ powders | Hydrothermal method | Solution of ammonium and iron sulphates in presence of hydrazine | Formation of oxide powders | -- | 7 |
| 2 | $\gamma$- $Fe_2O_3$ nanoparticles | Thermal decomposition | Iron nitride reacted with lauric acid | Kinetics of growth and vacancy ordering | -- | 8 |
| 3 | Core-shell nano-particles of $Fe_2O_3$ | Co-precipitation | Ferric chloride/HCl/$H_2O$ ferrous chloride hydrate/ NaOH mixture | Magnetic Properties | -- | 9 |
| 4 | Porous hollow micro- and nano- spheres | Calcination of nano-rods prepared through hydrothermal route | $\alpha$-FeOOH | Dielectric and Magnetic loss | Electromagnetic shielding | 10 |
| 5 | Hollow spheres of $Fe_2O_3$ | Hydrothermal method | Ferric nitrate, oxalic acid and urea | | Gas sensors | 11 |
| 6 | Large arrays of vertically aligned nano-wires | Heat treatment of iron substrates in oxidizing atmosphere | Iron substrates | High surface density | Nanodevices, catalysis | 12 |
| 7 | Thin Films and powders of $\alpha$-$Fe_2O_3$ | Sol-gel process followed by peptization of $\alpha$-FeOOH peptized by glacial acetic acid | Iron nitrate Hydrate, 2-metoxy-ethanol and acetylacetone mixture, Ferric chloride/ammonia, | Differences in Local Crystalline order | | 13,14 |
| 8 | Hollow spheres of $Fe_2O_3$ powders | Spray pyrolysis & heat treatment under reduction-oxidation conditions | Aqueous Iron nitrate aerosols | Decomposition efficiency of nano- crystalline particles | -- | 15 |
| 9 | Uniform sized $\gamma$- $Fe_2O_3$ particles of different degree of crystallinity | Single step combustion process by spray pyrolysis | Fe-acetyl acetone, Fe-ammonium citrate, Fe-nitrate dissolved in methanol | | Degree of crystallinity | 16 |
| 10 | Ferrite carbon composites | Pressure pyrolysis | Nickelocene and venyl ferrocene | Thermo-magnetic properties | -- | 17 |
| 11 | Nano-particles | Spray Pyrolysis | | | Spray Pyrolysis Mechanism | 18 |

18